\documentclass[submission,copyright,creativecommons,breaklinks]{eptcs}
\providecommand{\event}{AFL 2014}
\usepackage{breakurl}

\usepackage{amssymb}
\usepackage{amsfonts,amsmath}
\usepackage{amsthm}
\usepackage{latexsym,stmaryrd}
\usepackage[noend]{algorithmic}
\algsetup{indent=5.6mm}
\usepackage{color}
\usepackage{verbatim}
\usepackage{graphicx}
\usepackage{algorithm}
\usepackage
{hyperref}
\usepackage{breakurl}

\newcommand{\ifdraft}[1]{#1}
\definecolor{aocolour}{rgb}{0.7,0.8,1}
\definecolor{mbcolour}{rgb}{1,0.4,0.4}
\definecolor{reviewercolour}{rgb}{0.3,1,0.2}
\newcommand{\ao}[1]{\ifdraft{\noindent\colorbox{aocolour}{A.O.: #1}}}
\newcommand{\mb}[1]{\ifdraft{\noindent\colorbox{mbcolour}{M.B.: #1}}}
\newcommand{\review}[1]{\ifdraft{\noindent\colorbox{reviewercolour}{REVIEW: #1}}}

\makeatletter
\newcommand{\hacklabel}[1]{\protected@edef\@currentlabel{#1}}
\makeatother

{\begin{trivlist}\item[]\textbf{Lemma~\ref{#1}$'$.}\hacklabel{\ref{#1}$'$}}%
{\end{trivlist}}

\newcommand{\seepage}[1]{\marginpar{\scriptsize (p.~\pageref{#1})}}

\newcommand{\makeset}[2]{\{ \, #1 \mid #2 \, \}} 
\newcommand{\makesetbig}[2]{\big\{ \: #1 \; \big| \; #2 \: \big\}}

\renewcommand{\emptyset}{\varnothing}
\renewcommand{\epsilon}{\varepsilon}

\newcommand{\before}[1]{{\lhd} #1}
\newcommand{\beforeeq}[1]{{\trianglelefteqslant} #1}
\newcommand{\after}[1]{{\rhd} #1}
\newcommand{\aftereq}[1]{{\trianglerighteqslant} #1}

\newcommand{\e}{\varepsilon}
\renewcommand{\And}{\mathop{\&}}
\renewcommand{\phi}{\varphi}

\newcommand{\stt}[3]{#1 \langle #2 \rangle #3}
\newcommand{\cpt}[4]{#1\big(\stt{#2}{#3}{#4}\big)}

\newcommand{\len}[1]{\ensuremath{ |#1| }}

\newcommand{\NullableLeftEps}{\ensuremath{ \before{\epsilon}\text{-}\textsc{Nullable} }}
\newcommand{\NullableRightEps}{\ensuremath{ \after{\epsilon}\text{-}\textsc{Nullable} }}

\newcommand{\symbolbegin}{\ensuremath{ \underline{\mathrm{begin}} }} 
\newcommand{\symbolend}{\ensuremath{ \underline{\mathrm{end}} }} 

\newtheorem{definition}{Definition}
\newtheorem{theorem}{Theorem}
\newtheorem{lemma}{Lemma}

\numberwithin{claim}{theorem}
\theoremstyle{definition} \newtheorem{example}{Example}
\theoremstyle{definition} 
\newtheorem{construction}{Construction}
\theoremstyle{definition} 
\newtheorem{myalgorithm}{Algorithm}

\providecommand{\urlalt}[2]{\href{#1}{#2}}
\providecommand{\doi}[1]{DOI:~\urlalt{http://dx.doi.org/#1}{#1}}

\begin{document}

\sloppy

\title{Grammars with two-sided contexts\thanks{Supported by the Academy of Finland under grant 257857.}}
\def\titlerunning{Grammars with two-sided contexts}

\author{Mikhail Barash\institute{%
	Department of Mathematics and Statistics, University of Turku, Turku FI-20014, Finland}
	\institute{%
	Turku Centre for Computer Science,
	Turku FI-20520, Finland}
	\email{mikbar@utu.fi}
	\and
	Alexander Okhotin\institute{%
	Department of Mathematics and Statistics, University of Turku,
	Turku FI-20014, Finland}
	\email{alexander.okhotin@utu.fi}
}
\def\authorrunning{M. Barash, A. Okhotin}

\maketitle

\begin{abstract}
In a recent paper
(M. Barash, A. Okhotin, ``Defining contexts in context-free grammars'', LATA 2012),
the authors introduced an extension of the context-free grammars
equipped with an operator
for referring to the left context of the substring being defined.
This paper proposes a more general model,
in which context specifications may be two-sided,
that is, both the left and the right contexts
can be specified by the corresponding operators.
The paper gives the definitions
and establishes the basic theory
of such grammars,
leading to a normal form
and a parsing algorithm
working in time $\mathcal{O}(n^4)$, 
where $n$ is the length of the input string.
\end{abstract}


\section{Introduction}


The context-free grammars
are a logic for representing the syntax of languages,
in which the properties of longer strings
are defined by concatenating shorter strings
with known properties.
Disjunction of syntactic conditions
is represented in this logic
as multiple alternative rules
defining a single symbol.
One can further augment this logic
with conjunction and negation operations,
leading to \emph{conjunctive grammars}~\cite{Conjunctive}
and \emph{Boolean grammars}~\cite{BooleanGrammars}.
These grammars are context-free in the general sense of the word,
as they define the properties of each substring
independently of the context, in which it occurs.
Furthermore, most of the practically important features
of ordinary context-free grammars,
such as efficient parsing algorithms,
are preserved in their conjunctive and Boolean variants%
~\cite{BooleanGrammars,BooleanMatrix}. 
These grammar models
have been a subject of recent theoretical studies%
~\cite{AizikowitzKaminski_LR0,EsikKuich,Jez,Kountouriotis_et_al,Ziervogel}.

Not long ago, the authors~\cite{grammars_with_contexts_lata,grammars_with_contexts} 
proposed an extension of the context-free grammars
with special operators
for expressing the form of the \emph{left context},
in which the substring occurs.
For example, a rule $A \to BC \And \before{D}$
asserts that every string representable as $BC$
in a left context of the form described by $D$
therefore has the property $A$.
These grammars were motivated
by Chomsky's~\cite[p.~142]{Chomsky} well-known idea
of a phrase-structure rule
applicable only in some particular contexts.
Chomsky's own attempt to implement this idea
by string rewriting
resulted in a model equivalent to linear-space Turing machines,
in which the ``nonterminal symbols'',
meant to represent syntactic categories,
could be freely manipulated as tape symbols.
In spite of the name ``context-sensitive grammars'',
the resulting model was
unsuitable for describing the syntax of languages,
and thus failed to represent the idea
of a rule applicable in a context.

Taking a new start with this idea,
the authors~\cite{grammars_with_contexts}
defined \emph{grammars with one-sided contexts},
following the logical outlook on grammars,
featured in the work of Kowalski~\cite[Ch.~3]{Kowalski}
and of Pereira and Warren~\cite{PereiraWarren},
and later systematically developed by Rounds~\cite{Rounds}.
A grammar defines the truth value
of statements of the form
``a certain string has a certain property'',
and these statements
are deduced from each other
according to the rules of the grammar.
The resulting definition maintains
the underlying logic of the context-free grammars,
and many crucial properties of grammars are preserved:
grammars with one-sided contexts have parse trees,
can be transformed to a normal form
and have a cubic-time parsing algorithm~\cite{grammars_with_contexts}.
However, the model allowed specifying contexts only on one side,
and thus it implemented,
so to say, only one half of Chomsky's idea.

This paper continues the development
of formal grammars with context specifications
by allowing contexts in both directions.
The proposed \emph{grammars with two-sided contexts}
may contain such rules as
$A \to BC \And \before{D} \And \after{E}$,
which define any substring of the form $BC$
preceded by a substring of the form $D$
and followed by a substring of the form $E$.
If the grammar contains additional rules
$B \to b$, $C \to c$, $D \to d$ and $E \to e$,
then the above rule for $A$ asserts
that a substring $bc$ of a string $w=dbce$ has the property $A$.
However, this rule will not produce the same substring $bc$
occurring in another string $w'=dbcd$,
because its right context
does not satisfy the conjunct $\after{E}$.
Furthermore, the grammars allow expressing
the so-called \emph{extended right context} ($\aftereq{\alpha}$),
which defines the form of the current substring
concatenated with its right context,
as well as the symmetrically defined
\emph{extended left context} ($\beforeeq{\alpha}$).

In Section~\ref{section_deduction},
this intuitive definition
is formalized by deduction of propositions
of the form $\cpt{A}{u}{w}{v}$,
which states that
the substring $w$ occurring in the context between $u$ and $v$
has the property $A$,
where $A$ is a syntactic category defined by the grammar
(``nonterminal symbol'' in Chomsky's terminology).
Then, each rule of the grammar
becomes a schema for deduction rules,
and a string $w$ is generated by the grammar,
if there is a proof of the proposition $\cpt{S}{\epsilon}{w}{\epsilon}$.
A standard proof tree of such a deduction
constitutes a parse tree of the string $w$.

The next Section~\ref{section_examples}
presents basic examples
of grammars with two-sided contexts.
These examples model several types of cross-references,
such as declaration of identifiers before or after their use.

The paper then proceeds with developing
a normal form for these grammars,
which generalizes the Chomsky normal form
for ordinary context-free grammars.
In the normal form,
every rule is a conjunction of one or more \emph{base conjuncts}
describing the form of the current substring
(either as a concatenation of the form $BC$
or as a single symbol $a$),
with any context specifications
($\before{D}$, $\beforeeq{E}$, $\aftereq{F}$, $\after{H}$).
The transformation to the normal form,
presented in Section~\ref{section_normal_form},
proceeds in three steps.
First, all rules generating the empty string in any contexts
are eliminated.
Second, all rules with an explicit empty context specification
($\before{\epsilon}$, $\after{\epsilon}$)
are also eliminated.
The final step is elimination of any rules
of the form $A \to B \And \ldots$,
where the dependency of $A$ on $B$
potentially causes cycles in the definition.

Once the normal form is established,
a simple parsing algorithm for grammars with two-sided contexts
with the running time $\mathcal{O}(n^4)$
is presented in Section~\ref{section_parsing_algorithm}.
While this paper has been under preparation,
Rabkin~\cite{Rabkin} has developed
a more efficient and more sophisticated parsing algorithm
for grammars with two-sided contexts,
with the running time $\mathcal{O}(n^3)$.

\section{Definition} 
\label{section_deduction}

Ordinary context-free grammars
allow using the concatenation operation
to express the form of a string,
and disjunction to define alternative forms.
In conjunctive grammars,
the conjunction operation
may be used to assert that a substring being defined
must conform to several conditions at the same time.
The grammars studied in this paper
further allow operators
for expressing the form of the left context ($\before{}$, $\beforeeq{}$)
and the right context ($\after{}$, $\aftereq{}$)
of a substring being defined.

\begin{definition}
A grammar with two-sided contexts is a quadruple
$G=(\Sigma, N, R, S)$,
where
\begin{itemize}
\item	$\Sigma$ is the alphabet of the language being defined;
\item	$N$ is a finite set of auxiliary symbols
	(``nonterminal symbols'' in Chomsky's terminology),
	which denote the properties of strings defined in the grammar;
\item	$R$ is a finite set of grammar rules, each of the form
\begin{equation}
\label{eq:rule}
\begin{split}
A \to \alpha_1 \And \ldots \And \alpha_k \And
&\before{\beta_1} \And \ldots \And \before{\beta_m} \And
\beforeeq{\gamma_1} \And \ldots \And \beforeeq{\gamma_n} \And \\ \And
&\aftereq{\kappa_1} \And \ldots \And \aftereq{\kappa_{m'}} \And
\after{\delta_1} \And \ldots \And \after{\delta_{n'}},
\end{split}
\end{equation}
with $A \in N$, $k \geqslant 1$, $m,n,m',n' \geqslant 0$
and
$\alpha_i, \beta_i, \gamma_i, \kappa_i, \delta_i \in (\Sigma \cup N)^*$;
\item	$S \in N$ is a symbol
	representing 
	well-formed sentences of the language.
\end{itemize}
\end{definition}

If all rules in a grammar
have only left contexts
(that is, if $m'=n'=0$),
then this is a grammar with one-sided contexts~\cite{grammars_with_contexts}.
If no context operators are ever used ($m=n=m'=n'=0$),
this is a conjunctive grammar,
and if the conjunction is also never used ($k=1$),
this is an ordinary context-free grammar.

\begin{figure}[t]
%
%
\begin{center}
\includegraphics[scale=0.8]{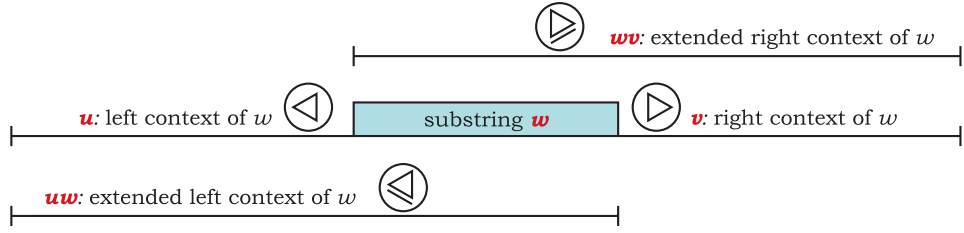}
\end{center}
\caption{A substring $w$ of a string $uwv$: four types of contexts.}
\label{f:substring_with_contexts}
\end{figure}

For each rule~(\ref{eq:rule}),
each term
$\alpha_i$, $\before{\beta_i}$, $\beforeeq{\gamma_i}$,
$\aftereq{\kappa_i}$ and $\after{\delta_i}$
is called a \emph{conjunct}.
Denote by $\stt{u}{w}{v}$ a substring $w \in \Sigma^*$,
which is preceded by $u \in \Sigma^*$
and followed by $v \in \Sigma^*$,
as illustrated in Figure~\ref{f:substring_with_contexts}.
Intuitively,
such a substring is generated by a rule~(\ref{eq:rule}),
if
\begin{itemize}
\item	each \emph{base conjunct} $\alpha_i=X_1 \ldots X_\ell$
	gives a representation of $w$
	as a concatenation of shorter substrings
	described by $X_1, \ldots, X_\ell$,
	as in context-free grammars;
\item	each conjunct $\before{\beta_i}$
	similarly describes the form
	of the \emph{left context} $u$;
\item	each conjunct $\beforeeq{\gamma_i}$
	describes the form of the
	\emph{extended left context} $uw$;
\item	each conjunct $\aftereq{\kappa_i}$
	describes the \emph{extended right context} $wv$;
\item	each conjunct $\after{\delta_i}$
	describes the \emph{right context} $v$.
\end{itemize}

The semantics of grammars with two-sided contexts
are defined by a deduction system 
of elementary propositions (items)
of the form ``a string $w \in \Sigma^*$
written in a left context $u \in \Sigma^*$
and in a right context $v \in \Sigma^*$
has the property $X \in \Sigma \cup N$'',
denoted by $\cpt{X}{u}{w}{v}$.
The deduction begins with axioms:
any symbol $a \in \Sigma$ written in any context
has the property $a$,
denoted by $\cpt{a}{u}{a}{v}$ for all $u,v \in \Sigma^*$.
Each rule in $R$
is then regarded
as a schema for deduction rules.
For example, a rule $A \to BC$
allows making deductions of the form
\begin{align*}
	\cpt{B}{u}{w}{w'v},
	\cpt{C}{uw}{w'}{v}
	\vdash_G
	\cpt{A}{u}{ww'}{v}
	&& (\text{for all $u,w,w',v \in \Sigma^*$}),
\end{align*}
which is essentially a concatenation of $w$ and $w'$
that respects the contexts.
If the rule is of the form $A \to BC \And \before{D}$,
this deduction requires an extra premise:
\begin{align*}
	\cpt{B}{u}{w}{w'v},
	\cpt{C}{uw}{w'}{v},
	\cpt{D}{\epsilon}{u}{ww'v}
	&\vdash_G
	\cpt{A}{u}{ww'}{v}.
\intertext{And if the rule is $A \to BC \And \aftereq{F}$,
the deduction proceeds as follows:}
	\cpt{B}{u}{w}{w'v},
	\cpt{C}{uw}{w'}{v},
	\cpt{F}{u}{ww'v}{\epsilon}
	&\vdash_G
	\cpt{A}{u}{ww'}{v}.
\end{align*}
The general form of deduction schemata
induced by a rule in $R$
is defined below.

\begin{definition}
\label{def:deduction_system}
Let $G=(\Sigma, N, R, S)$ be a grammar with two-sided contexts.
Define the following deduction system
of items of the form
$\cpt{X}{u}{w}{v}$, with $X \in \Sigma \cup N$ and $u,w,v \in \Sigma^*$.
There is a single axiom scheme
$\vdash_G \cpt{a}{u}{a}{v}$,
for all $a \in \Sigma$ and $u,v \in \Sigma^*$.
Each rule
(\ref{eq:rule})
in $R$ 
defines the following scheme for deduction rules:
\begin{equation*}
	I \vdash_G \cpt{A}{u}{w}{v},
\end{equation*}
for all $u, w, v \in \Sigma^*$
and for every set of items $I$ satisfying the below properties:
\begin{itemize}
\item	For every base conjunct $\alpha_i=X_1 \ldots X_\ell$,
	with $\ell \geqslant 0$ and $X_j \in \Sigma \cup N$,
	there should exist a partition $w=w_1 \ldots w_\ell$
	with $\cpt{X_j}{u w_1 \ldots w_{j-1}}{w_j}{w_{j+1} \ldots w_\ell v} \in I$
	for all $j \in \{1, \ldots, \ell\}$.
\item	For every conjunct $\before{\beta_i}=\before{X_1 \ldots X_\ell}$
	there should be such a partition $u=u_1 \ldots u_\ell$,
	that $\cpt{X_j}{u_1 \ldots u_{j-1}}{u_j}{u_{j+1} \ldots u_\ell w v} \in I$
	for all $j \in \{1, \ldots, \ell\}$.
\item	Every conjunct $\beforeeq{\gamma_i}=\beforeeq{X_1 \ldots X_\ell}$
	should have a corresponding partition $uw=x_1 \ldots x_\ell$
	with $\cpt{X_j}{x_1 \ldots x_{j-1}}{x_j}{x_{j+1} \ldots x_\ell v} \in I$
	for all $j \in \{1, \ldots, \ell\}$.
\item	For every conjunct $\after{\delta_i}$ and $\aftereq{\kappa_i}$,
	the condition is defined symmetrically.
\end{itemize}
Then the language generated by a symbol $A \in N$
is defined as
\begin{equation*}
L_G(A) = \makeset{\stt{u}{w}{v}}{u,w,v \in \Sigma^*, \: \vdash_G \cpt{A}{u}{w}{v}}.
\end{equation*}
The language generated by the grammar $G$
is the set of all strings with empty left and right contexts
generated by $S$:
$L(G)=\makeset{w}{w \in \Sigma^{*}, \: \vdash_G \cpt{S}{\epsilon}{w}{\epsilon}}$.
\end{definition}

The following trivial example of a grammar
is given to illustrate the definitions.

\begin{example}\label{example:abca_grammar}
Consider the grammar with two-sided contexts
that defines the singleton language $\{abca\}$:
\begin{eqnarray*}
S &\to& a S \ | \ S a \ | \ B C \\
A &\to& a \\
B &\to& b \And \before{A} \\
C &\to& c \And \after{A}
\end{eqnarray*}
\end{example}

The deduction given below proves
that the string $abca$ has the property $S$.
\begin{align*}
	& \vdash \cpt{a}{\epsilon}{a}{bca}
		&& (axiom) \\
	& \vdash \cpt{b}{a}{b}{ca}
		&& (axiom) \\
	& \vdash \cpt{c}{ab}{c}{a}
		&& (axiom) \\
	& \vdash \cpt{a}{abc}{a}{\epsilon}
		&& (axiom) \\
	\cpt{a}{\epsilon}{a}{bca}
	& \vdash \cpt{A}{\epsilon}{a}{bca}
		&& (A \to a) \\
	\cpt{b}{a}{b}{ca}, \cpt{A}{\epsilon}{a}{bca}
	& \vdash \cpt{B}{a}{b}{ca}
		&& (B \to b \And \before{A}) \\
	\cpt{a}{abc}{a}{\epsilon}
	& \vdash \cpt{A}{abc}{a}{\epsilon}
		&& (A \to a) \\
	\cpt{c}{ab}{c}{a}, \cpt{A}{abc}{a}{\epsilon}
	& \vdash \cpt{C}{ab}{c}{a}
		&& (C \to c \And \after{A}) \\
	\cpt{B}{a}{b}{ca}, \cpt{C}{ab}{c}{a}
	& \vdash \cpt{S}{a}{bc}{a}
		&& (S \to BC) \\
	\cpt{a}{\epsilon}{a}{bca}, \cpt{S}{a}{bc}{a}
	& \vdash \cpt{S}{\epsilon}{abc}{a}
		&& (S \to a S) \\
	\cpt{S}{\epsilon}{abc}{a}, \cpt{a}{abc}{a}{\epsilon} 
	& \vdash \cpt{S}{\epsilon}{abca}{\epsilon}
		&& (S \to S a)
\end{align*}

\begin{figure}[t]
%
%
\begin{center}
\includegraphics[scale=0.8]{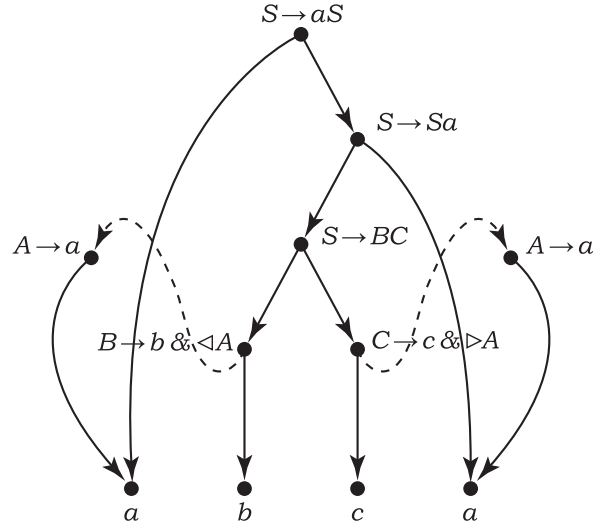}
\end{center}
\caption{A parse tree of the string $abca$
	according to the grammar in Example~\ref{example:abca_grammar}.}
\label{f:abca_tree}
\end{figure}

Another possible definition of grammars with contexts
is by directly expressing them
in first-order logic over positions in a string~\cite{Rounds}.
Nonterminal symbols become \emph{binary predicates},
with the arguments referring to positions in the string.
Each predicate $A(x, y)$
is defined by a formula $\varphi_A(x, y)$
that states the condition
of a substring delimilited by positions $x$ and $y$
having the property $A$.
There are built-in unary predicates
$a(x)$, for each $a \in \Sigma$,
which assert that the symbol in position $x$ in the string is $a$,
and binary predicates $x<y$ and $x=y$ for comparing positions.
Arguments to predicates
are given as \emph{terms},
which are either variables ($t=x$)
or constants referring to the first and the last positions
($t=\symbolbegin$, $t=\symbolend$),
and which may be incremented ($t+1$) or decremented ($t-1$).
Each formula is constructed from predicates
using conjunction,
disjunction
and first-order existential quantification.

\begin{example}
The grammar from Example~\ref{example:abca_grammar}
is expressed by the following formulae
defining predicates
$S(x,y)$, $B(x,y)$, $A(x,y)$ and $C(x,y)$.
\begin{eqnarray*}
S(x,y) &=&
	\left( a(x) \land S(x+1,y) \right)
	\,\lor\,
	\left( S(x,y-1) \land a(y) \right)
	\,\lor\,
	\left( \exists z \left( x < z < y 
	\land B(x,z) \land C(z,y) \right) \right) \\
A(x,y) &=&
	a(x) \land x+1 = y \\
B(x,y) &=&
	b(x) \land x+1 = y \land A(\symbolbegin, x) \\
C(x,y) &=&
	c(x) \land x+1 = y \land A(y, \symbolend)
\end{eqnarray*}
The membership of a string $w$ is expressed
by the statement $S(\symbolbegin, \symbolend)$,
which may be true of false.
\end{example}

\section{Examples}\label{section_examples}

This section presents several examples
of grammars with two-sided contexts
generating important syntactic constructs.
All examples use ordinary context-free elements,
such as a grammar for $\makeset{a^nb^n}{n \geqslant 0}$,
and combine these elements using the new context operators.
This leads to natural specifications of languages
in the style of classical formal grammars.

Consider the problem
of checking declaration of identifiers before their use:
this construct can be found in all kinds of languages,
and it can be expressed by a conjunctive grammar~\cite[Ex.~3]{BooleanSurvey}.
The variant of this problem,
in which the identifiers may be declared \emph{before or after} their use,
is also fairly common:
consider, for instance, the declaration of classes in C++,
where an earlier defined method
can refer to a class member defined later.
However, no conjunctive grammar
expressing this construct is known.

A grammar with one-sided contexts for declarations before or after use
has recently been constructed by the authors~\cite{grammars_with_contexts}.
That grammar used context specifications,
along with iterated conjunction,
to express what would be more naturally expressed
in terms of two-sided contexts.
In the model proposed in this paper,
the same language can be defined in a much more natural way.

\begin{example}[cf.~grammar with one-sided contexts~{\cite[Ex.~4]{grammars_with_contexts}}]
\label{declarations_before_or_after_use_example}
Consider the language
\begin{equation*}\label{declarations_before_or_after_use_language}
	\makeset{u_1 \ldots u_n}{\text{for every $u_i$,
		\textbf{either}
		$u_i \in a^* c$,
		\textbf{or}
		$u_i = b^k c$ and
		there exists $j \in \{1, \ldots, n\}$
		with $u_j=a^k c$}}.
\end{equation*}
Substrings of the form $a^k c$ represent declarations,
while every substring of the form $b^k c$ is a reference
to a declaration of the form $a^k c$.

This language is generated by the following grammar.
\begin{equation*}\begin{array}{rcl@{\quad \qquad}rcl}
S &\to& AS \ | \ CS \ | \ DS \ | \ \epsilon
	& C &\to& B \And \beforeeq{EFc} \\
A &\to& a A \ | \ c
	& D &\to& B \And \aftereq{HcE} \\
B &\to& b B \ | \ c
	& F &\to& a F b \ | \ c E \\
E &\to& AE \ | \ BE \ | \ \epsilon
	& H &\to& b H a \ | \ c E
\end{array}\end{equation*}

The idea of the grammar
is that $S$ should generate a substring
$\stt{u_1 \ldots u_\ell}{u_{\ell+1} \ldots u_n}{\epsilon}$,
with $0 \leqslant \ell \leqslant n$
and $u_i \in a^* c \cup b^* c$,
if and only if
every reference in $u_{\ell+1} \ldots u_n$
has a corresponding declaration
somewhere in the whole string $u_1 \ldots u_n$.
The rules for $S$ define all substrings satisfying this condition
inductively on their length,
until the entire string
$\stt{\epsilon}{u_1 \ldots u_n}{\epsilon}$ is defined.
The rule $S \to \epsilon$ defines the base case:
the string
$\stt{u_1 \ldots u_n}{\epsilon}{\epsilon}$
has the desired property.
The rule $S \to CS$
appends a reference of the form $b^*c$,
restricted by an extended left context $\beforeeq{EFc}$,
which ensures that this reference
has a matching \emph{earlier} declaration;
here $E$ represents the prefix of the string
up to that earlier declaration,
while $F$ matches the symbols $a$ in the declaration
to the symbols $b$ in the reference.
The possibility of a \emph{later} declaration
is checked by another rule $S \to DS$,
which adds a reference of the form $b^*c$
with an extended right context $\aftereq{HcE}$,
where $H$ is used to match the $b$s forming this reference
to the $a$s in the later declaration.
\end{example}


The next example abstracts another syntactic mechanism---%
\emph{function prototypes}---%
found in the C programming language
and,
under the name of \emph{forward declarations},
in the programming language Pascal.


\begin{example}\label{example:prototypes}
Consider the language
\begin{subequations}
\begin{align}
\label{eq:prototype_declarations__prototype}
\big\{
u_1 \ldots u_n
\: \big| \:
\text{for every $u_i$}, \;
	& \text{\textbf{either} $u_i = a^k c$ and there exists $j > i$,
		such that $u_j = d^k c$,} \\
\label{eq:prototype_declarations__reference}
	& \text{\textbf{or} $u_i = b^k c$ and there exists $j < i$,
		for which $u_j = a^k c$}\big\}.
\end{align}
\end{subequations}

A substring of the form $a^k c$
represents a function prototype
and a substring $d^k c$
represents its body.
Calls to functions are expressed as substrings $b^k c$.
Condition~(\ref{eq:prototype_declarations__prototype})
means that every prototype must be followed by its body,
and restriction~(\ref{eq:prototype_declarations__reference})
requires that references are only allowed to declared prototypes.

This language can be generated by the following
grammar with two-sided contexts.
\begin{equation*}\begin{array}{rcl@{\quad\qquad}rcl@{\quad\qquad}rcl}
S &\to& U S \ | \ V S \ | \ D S \ | \ \epsilon
	& D &\to& d D \ | \ c
		& E &\to& A E \ | \ B E \ | \ D E \ | \ \epsilon\\
A &\to& a A \ | \ c
	& U &\to& A \And \aftereq{H c E}
		& H &\to& a H d \ | \ c E\\
B &\to& b B \ | \ c
	& V &\to& B \And \beforeeq{E F c}
		& F &\to& a F b \ | \ c E\\
\end{array}\end{equation*}

The rules $S \to US$ and $U \to A \And \aftereq{HcE}$
append a prototype $a^k c$
and the extended right context of the form
$a^k c \ldots d^k c \ldots$
ensures that this prototype
has a matching body somewhere \emph{later}
within the string.
The rules $S \to VS$ and $V \to B \And \beforeeq{EFc}$
append a reference $b^k c$,
and the context specification $\ldots a^k c \ldots b^k c$
checks that it has a matching prototype \emph{ealier}
in the string.
Function bodies $d^k c$
are added by the rule $S \to DS$.
Using these rules,
$S$ generates substrings of the form
$\stt{u_1 \ldots u_\ell}{u_{\ell+1} \ldots u_n}{\epsilon}$,
with $0 \leqslant \ell \leqslant n$ and
$u_i \in a^*c \cup b^*c \cup d^*c$,
such that
every prototype $u_i = a^k c$ in $u_{\ell+1} \ldots u_n$
has a corresponding body $d^k c$ in $u_{i+1} \ldots u_n$
and
every reference $u_i = b^k c$ in $u_{\ell+1} \ldots u_n$
has a corresponding prototype $a^k c$ in $u_1 \ldots u_{i-1}$.

\end{example}

The next example gives a grammar with contexts
that defines reachability on graphs.
Sudborough~\cite{Sudborough} defined a linear context-free grammar
for a special encoding
of the graph reachability problem
on acyclic graphs,
in which every arc goes from a lower-numbered vertex
to a higher-numbered vertex.
The grammar presented below allows any graphs
and uses a direct encoding.
This example
illustrates the ability of grammars with contexts
to define various kinds of cross-references.

\begin{example}\label{graph_example}
Consider encodings of directed graphs
as strings of the form
$b^s \:
a^{i_1} b^{j_1} \: a^{i_2} b^{j_2} \: \ldots \: a^{i_n} b^{j_n} \:
a^t$,
with $s, t \geqslant 1, \: n \geqslant 0, \: i_k,j_k \geqslant 1$,
where each block $a^i b^j$
denotes an arc
from vertex number $i$ to vertex number $j$,
while the prefix $b^s$ and the suffix $a^t$
mark $s$ as the source vertex and $t$ as the target.
Then the following grammar
defines all graphs with a path from $s$ to $t$.
\begin{equation*}\begin{array}{rcl@{\quad \qquad}rcl}
S &\to& F D C A \ | \ F 
	& &&\\
A &\to& a A \ | \ c
	& D &\to& B \And \beforeeq{B C E} \ | \ B \And \aftereq{FDCA} \ | \ B \And \aftereq{F} \\
B &\to& b B \ | \ c
	& E &\to& a E b \ | \ D C A \\
C &\to& A B C \ | \ \epsilon
	& F &\to& b F a \ | \ b C a
\end{array}\end{equation*}
\end{example}

The grammar is centered around the nonterminal $D$,
which generates all such substrings
$\stt{b^s a^{i_1} b^{j_1} \ldots a^{i_k}}{b^{j_k}}{a^{i_{k+1}}b^{j_{k+1}} \ldots a^{i_n} b^{j_n} a^t}$
that there is a path from $j_k$ to $t$ in the graph.
If this path is empty, then $j_k = t$.
Otherwise, the first arc in the path
can be listed either to the left or to the right
of $b^k$.
These three cases are handled by the three rules for $D$.
Each of these rules generates $b^{j_k}$ by the base conjunct $B$,
and then uses an extended left or right context operator
to match $b^{j_k}$ to the tail of the next arc or to $a^t$.

The rule $D \to B \And \beforeeq{B C E}$
considers the case when the next arc
in the path is located to the left of $b^{j_k}$.
Let this arc be $a^{i_\ell}b^{j_\ell}$, for some $\ell<k$.
Then the extended left context $BCE$
covers the substring
$b^s a^{i_1}b^{j_1} \ldots a^{i_\ell}b^{j_\ell} \ldots a^{i_k}b^{j_k}$.
The concatenation $BC$
skips the prefix $b^s a^{i_1}b^{j_1} \ldots a^{i_{\ell-1}}b^{j_{\ell-1}}$,
and then the nonterminal $E$
matches $a^{i_\ell}$ to $b^{j_k}$,
verifying that $i_\ell=j_k$.
After this, the rule $E \to DCA$
ensures that the substring $b^{j_\ell}$ is generated by $D$,
that is, that there is a path from $j_\ell$ to $t$.
The concatenation $CA$
skips the inner substring $a^{i_{\ell+1}}b^{j_{\ell+1}} \ldots a^{i_k}$.

The second rule $D \to B \And \aftereq{FDCA}$
searches for the next arc to the right of $b^{j_k}$.
Let this be an $\ell$-th arc in the list, with $\ell>k$.
The extended right context $FDCA$
should generate the suffix
$b^{j_k} \ldots a^{i_\ell}b^{j_\ell} \ldots a^{i_n}b^{j_n} a^t$.
The symbol $F$ covers the substring $b^{j_k} \ldots a^{i_\ell}$,
matching $b^{j_k}$ to $a^{i_\ell}$.
Then, $D$ generates the substring $b^{j_\ell}$,
checking that there is a path from $j_\ell$ to $t$.
The concatenation $CA$ skips the rest of the suffix.

Finally, if the path is of length zero,
that is, $j_k=t$,
then the rule $D \to B \And \aftereq{F}$
uses $F$ to match $b^{j_k}$
to the suffix $a^t$ in the end of the string.

Once the symbol $D$ checks the path from any vertex to the vertex $t$,
for the initial symbol $S$,
it is sufficient to match $b^s$ in the beginning of the string
to any arc $a^{j_k} b^{j_k}$, with $j_k=s$.
This is done by the rule $S \to FDCA$,
which operates in the same way as the second rule for $D$.
The case of $s$ and $t$ being the same node
is handled by the rule $S \to F$.

All the above examples use identifiers given in unary,
which are matched by rules of the same kind
as the rules defining the language $\makeset{a^n b^n}{n \geqslant 0}$.
These examples can be extended
to use identifiers over an arbitrary alphabet $\Sigma$,
owing to the fact
that there is a conjunctive grammar
generating the language $\makeset{w\#w}{w \in \Sigma^*}$,
for some separator $\# \notin \Sigma$~\cite{Conjunctive,BooleanSurvey}.


\section{Normal form}\label{section_normal_form}

An ordinary context-free grammar
can be transformed to the Chomsky normal form,
with the rules restricted to $A \to BC$ and $A \to a$,
with $B,C \in N$ and $a \in \Sigma$.
This form has the following generalization
to grammars with contexts.

\begin{definition}
A grammar with two-sided contexts $G=(\Sigma,N,R,S)$
is said to be in the binary normal form,
if each rule in $R$ is of one of the forms
\begin{align*}
	A &\to B_1 C_1 \And \ldots \And B_k C_k \And
		\before{D_1} \And \ldots \And \before{D_m} \And
		\beforeeq{E_1} \And \ldots \And \beforeeq{E_n} \And
		\aftereq{F_1} \And \ldots \And \aftereq{F_{n'}} \And
		\after{H_1} \And \ldots \And \after{H_{m'}}, \\
	A &\to a \And 
		\before{D_1} \And \ldots \And \before{D_m} \And
		\beforeeq{E_1} \And \ldots \And \beforeeq{E_n} \And
		\aftereq{F_1} \And \ldots \And \aftereq{F_{n'}} \And
		\after{H_1} \And \ldots \And \after{H_{m'}},
\end{align*}
where $k \geqslant 1$, $m,n,n',m' \geqslant 0$,
$B_i, C_i, D_i, E_i, F_i, H_i \in N$, $a \in \Sigma$.
\end{definition}

The transformation to the normal form
consists of three stages:
first, removing all \emph{empty conjuncts} $\epsilon$;
secondly, eliminating \emph{empty contexts}
($\before{\epsilon}$, $\after{\epsilon}$);
finally, getting rid of \emph{unit conjuncts}
of the form $B$, with $B \in N$.


The first step is
the removal of all rules of the form
$A \to \epsilon \And \ldots$,
so that no symbols generate $\epsilon$,
while all non-empty strings are generated as before.
As generation of longer strings
may depend on the generation of $\epsilon$,
already for ordinary context-free grammars,
such a transformation requires adding extra rules
that simulate the same dependence
without actually generating any empty strings.

\begin{example}\label{example:nullable_context_free}
Consider the following context-free grammar,
which defines the language
$\{abc, ab, ac, a, bcd, bd, cd, d\}$.
\begin{eqnarray*}
S &\to& a A \ | \ A d \\
A &\to& BC\\
B &\to& \epsilon \ | \ b \\
C &\to& \epsilon \ | \ c
\end{eqnarray*}
Since $B$ generates the empty string,
the rule $A \to BC$ can be used to generate just $C$;
therefore, once the rule $B \to \epsilon$ is removed,
one should add a new rule $A \to C$,
in which $B$ is omitted.
Similarly one can remove the rule $C \to \epsilon$
and add a ``compensatory'' rule $A \to B$.
Since both $B$ and $C$ generate $\epsilon$,
so does $A$ by the rule $A \to BC$.
Hence,
extra rules $S \to a$ and $S \to d$,
where $A$ is omitted,
have to be added.
\end{example}

An algorithm for carrying out such a transformation
first calculates the set of nonterminals that generate the empty string,
known as $\textsc{Nullable}(G) \subseteq N$,
and then uses it to reconstruct the rules of the grammar.

This set is calculated as a least upper bound
of an ascending sequence of sets $\textsc{Nullable}_i(G)$.
The set
$\textsc{Nullable}_1(G) = \makeset{A \in N}{A \to \epsilon \in R}$
contains all nonterminals which directly define
the empty string.
Every next set
$\textsc{Nullable}_{i+1}(G) = \makeset{A \in N}{A \to \alpha \in R, \: \alpha \in \textsc{Nullable}_i^*(G)}$
contains nonterminals that generate $\epsilon$
by the rules referring to other nullable nonterminals.
This knowledge is given by the Kleene star
of $\textsc{Nullable}_i(G)$.

For the grammar in Example~\ref{example:nullable_context_free},
the calculation of the set $\textsc{Nullable}(G)$
proceeds as follows:
\begin{eqnarray*}
\textsc{Nullable}_0(G) &=& \emptyset,\\
\textsc{Nullable}_1(G) &=& \big\{ B, C \big\},\\
\textsc{Nullable}_2(G) &=& \big\{ B, C, A \big\},
\end{eqnarray*}
and $\textsc{Nullable}(G) = \textsc{Nullable}_2(G)$.

The same idea works for conjunctive grammars as well~\cite{Conjunctive}.
For grammars with contexts~\cite{grammars_with_contexts},
the generation of the empty string additionally depends on the left contexts,
in which the string occurs.
This requires an elaborated version of the set $\textsc{Nullable}(G)$,
formed of nonterminals
along with the information about the left contexts
in which they may define $\epsilon$.

In order to eliminate null conjuncts
in case of grammars with two-sided contexts,
one has to consider
yet another
variant of the set $\textsc{Nullable}(G)$,
which respects both left and right contexts.

\begin{example}\label{example:nullable_grammar_with_two_sided_contexts}
Consider the following grammar with two-sided contexts,
obtained by adding context restrictions
to the grammar in Example~\ref{example:nullable_context_free};
this grammar defines the language
$L = \{abc, ac, bcd, bd\}$.
\begin{eqnarray*}
S &\to& a A \ | \ A d\\
A &\to& B C\\
B &\to& \epsilon \And \before{D} \ | \ b\\
C &\to& \epsilon \And \after{E} \ | \ c\\
D &\to& a\\
E &\to& d
\end{eqnarray*}
In this grammar, the nonterminal $B$ generates
the empty string only in a left context of the form
defined by $D$,
while $C$ defines the empty string only in a right context
of the form $E$.
In those contexts where \emph{both} $B$ and $C$ generate $\epsilon$,
so can $A$, by the rule $A \to BC$.
\end{example}

The information about the left and right contexts,
in which a nonterminal generates the empty string,
is to be stored in
the set $\textsc{Nullable}(G)$,
which is defined as a subset of $2^N \times N \times 2^N$.
An element $(U, A, V)$ of this set represents
an intuitive idea that $A$ defines $\epsilon$
in a left context of the form described by each nonterminal in $U$,
and in a right context of the form given by nonterminals in $V$.

For the grammar in Example~\ref{example:nullable_grammar_with_two_sided_contexts},
such a set $\textsc{Nullable}(G)$
is constructed as follows.
\begin{eqnarray*}
\textsc{Nullable}_0(G) &=& \emptyset \\
\textsc{Nullable}_1(G) &=& \big\{ (\{D\}, B, \emptyset), (\emptyset, C, \{E\}) \big\} \\
\textsc{Nullable}_2(G) &=& \big\{ (\{D\}, B, \emptyset), (\emptyset, C, \{E\}), (\{D\}, A, \{E\}) \big\}
\end{eqnarray*}
Then $\textsc{Nullable}(G) = \textsc{Nullable}_2(G)$.
The elements $(\{D\}, B, \emptyset)$ and
$(\emptyset, C, \{E\})$ are obtained
directly from the rules of the grammar,
and the element $(\{D\}, A, \{E\})$
represents the ``concatenation'' $BC$
in the rule for $A$.
Note the similarity of this construction
to the one for the ordinary grammar in Example~\ref{example:nullable_context_free}:
the construction given here
is different only in recording information about the contexts.

The above ``concatenation'' of triples
$(\{D\}, B, \emptyset)$ and $(\emptyset, C, \{E\})$
should be defined to accumulate both left and right contexts.
This can be regarded
as a generalization of the Kleene star
to sets of triples,
denoted by $\textsc{Nullable}^\star(G)$.
Formally,
$\textsc{Nullable}^\star(G)$
is the set of all triples
$(U_1 \cup \ldots \cup U_\ell, \: A_1 \ldots A_\ell, \: V_1 \cup \ldots \cup V_\ell)$
with $\ell \geqslant 0$
and $(U_i, A_i, V_i) \in \textsc{Nullable}(G)$.
The symbols $A_i$
are concatenated,
while their left and right contexts are accumulated.
In the special case when $\ell = 0$,
the concatenation of zero symbols is the empty string,
and thus $\emptyset^\star = \big\{ (\emptyset, \epsilon, \emptyset) \big\}$.

Before giving a formal definition
of the set $\textsc{Nullable}(G)$,
assume, for the sake of simplicity,
that context operators are only applied
to single nonterminal symbols,
that is, every rule is of the form
\begin{equation}
\label{eq:rule_alpha_D_E_F_H}
\begin{split}
A \to \alpha_1 \And \ldots \And \alpha_k \And
\before{D_1} \And \ldots \And \before{D_m} \And
\beforeeq{E_1} \And \ldots \And \beforeeq{E_n} \And
\aftereq{F_1} \And \ldots \And \aftereq{F_{m'}} \And
\after{H_1} \And \ldots \And \after{H_{n'}},
\end{split}
\end{equation}
with $A \in N$, $k \geqslant 1$, $m,n,m',n' \geqslant 0$,
$\alpha_i \in (\Sigma \cup N)^*$
and $D_i, E_i, F_i, H_i \in N$.
As will be shown in Lemma~\ref{lemma:almost_normal_form},
there is no loss of generality
in this assumption.

\begin{definition}
\label{def:Nullable}
Let $G = (\Sigma, N, R, S)$ be a grammar with two-sided contexts
with all rules of the form~(\ref{eq:rule_alpha_D_E_F_H}).
Construct the sequence of sets
$\textsc{Nullable}_i(G) \subseteq 2^N \times N \times 2^N$,
for $i \geqslant 0$,
as follows.

Let $\textsc{Nullable}_0(G) = \emptyset$.
Every next set
$\textsc{Nullable}_{i+1}(G)$
contains the following triples:
for every rule~(\ref{eq:rule_alpha_D_E_F_H})
and for every $k$ triples
$(U_1, \alpha_1, V_1)$, \ldots, $(U_k, \alpha_k, V_k)$
in $\textsc{Nullable}_i^\star(G)$,
the triple
	$\big(
	\{D_1,\ldots,D_m, E_1, \ldots, E_n\}
	\cup \{U_1, \ldots, U_k\},
	\; A, \;
	\{F_1,\ldots,F_{m'}, H_1, \ldots, H_{n'}\}
	\cup \{V_1, \ldots, V_k\}
	\big)$
is in $\textsc{Nullable}_{i+1}(G)$.

Finally, let
$\textsc{Nullable}(G) = \bigcup_{i \geqslant 0} \textsc{Nullable}_{i}(G)$.
\end{definition}

The next lemma explains how exactly
the set $\textsc{Nullable}(G)$
represents the generation of the empty string
by different nonterminals in different contexts.

\begin{lemma}\label{lemma:nullable}
Let $G = (\Sigma,N,R,S)$ be a grammar with contexts,
let $A \in N$ and $u, v \in \Sigma^*$.
Then, $\stt{u}{\epsilon}{v} \in L_G(A)$
	if and only if
there is a triple
$(\{J_1, \ldots, J_s\}, A, \{K_1, \ldots, K_t\})$ in $\textsc{Nullable}(G)$,
such that $\stt{\epsilon}{u}{v} \in L_G(J_i)$ for all $i$
and $\stt{u}{v}{\epsilon} \in L_G(K_j)$ for all $j$.
\end{lemma}

The plan is to reconstruct the grammar,
so that for every triple
$(\{J_1, \ldots, J_s\}, A, \{K_1, \ldots, K_t\})$ in $\textsc{Nullable}(G)$,
and for every occurrence of $A$
in the right-hand side of any rule,
the new grammar contains a companion rule,
in which $A$ is omitted
and context operators for $J_i$ and $K_i$
are introduced.

The following case requires special handling in the new grammar.
Assume that $A$ generates $\epsilon$ in the empty left context
(that is, $u=\epsilon$ in Lemma~\ref{lemma:nullable}).
This is reflected by a triple 
$(\{J_1, \ldots, J_s\}, A, \{K_1, \ldots, K_t\})$ in $\textsc{Nullable}(G)$,
in which all symbols $J_i$
also generate $\epsilon$ in the left context $\epsilon$.
The latter generation
may in turn involve some further right context operators.
In the new grammar,
the left context will be explicitly set to be empty ($\before{\epsilon}$),
whereas all those right contexts should be assembled
together with the set $\{K_1, \ldots, K_t\}$,
and used in the new rules, where $A$ is omitted.
This calculation of right contexts
is done in the following special variant of the set $\textsc{Nullable}$.

\begin{definition}\label{def:left_Nullable}
Let $G = (\Sigma, N, R, S)$ be a grammar.
Define sets
$\NullableLeftEps_i(G) \subseteq N \times 2^N$,
with $i \geqslant 0$:
\begin{align*}
	\NullableLeftEps_0(G) &= \makeset{(A, V)}{(\emptyset, A, V) \in \textsc{Nullable}(G)},
		\\
	\NullableLeftEps_{i+1}(G) &=
	\big\{
(A, V \cup V_1 \cup \ldots \cup V_s)
\: \big| \:
(\{J_1, \ldots, J_s\}, A, V) \in \textsc{Nullable}(G), \\
&\hspace*{4cm}
\exists \: V_1, \ldots, V_s \subseteq N:
(J_i, V_i) \in \NullableLeftEps_i(G) \big\}.
\end{align*}
Let
$\NullableLeftEps(G) = \bigcup_{i \geqslant 0} \NullableLeftEps_i(G)$.
\end{definition}

\begin{lemma}\label{lemma:left_nullable}
Let $G = (\Sigma, N, R, S)$ be a grammar,
let $A \in N$ and $v \in \Sigma^*$.
Then
$\stt{\epsilon}{\epsilon}{v} \in L_G(A)$
if and only if
there is a pair
$(A, \{K_1, \ldots, K_t\})$ in $\NullableLeftEps(G)$,
such that
$\stt{\epsilon}{v}{\epsilon} \in L_G(K_i)$ for all $i$.
\end{lemma}

There is a symmetrically defined set
$\NullableRightEps(G) \subseteq 2^N \times N$,
which characterizes the generation of $\epsilon$
in an empty right context.

With the generation of the empty string
represented in these three sets,
a grammar with two-sided contexts
is transformed to the normal form
as follows.
First, it is convenient
to simplify the rules of the grammar,
so that every concatenation is of the form $BC$, with $B, C \in N$,
and the context operators are only applied to individual nonterminals.
For this, base conjuncts $\alpha$ with $\len{\alpha} > 2$
and context operators
$\before{\alpha}$,
$\beforeeq{\alpha}$,
$\aftereq{\alpha}$ and
$\after{\alpha}$
with $\len{\alpha} > 1$
are shortened by introducing new nonterminals.

\begin{lemma}\label{lemma:almost_normal_form}
For every grammar $G_0 = (\Sigma, N_0, R_0, S_0)$,
there exists and can be effectively
constructed another grammar $G = (\Sigma, N, R, S)$
generating the same language,
with all rules of the form:
\begin{subequations}\label{eq:rules_almost_normal_form}
\begin{align}
\label{eq:almost_a} A &\to a\\
\label{eq:almost_BC} A &\to BC\\
\label{eq:almost_B_D_E_F_H}
	A &\to 	B_1 \And \ldots \And B_k \And 
			\before{D_1} \And \ldots \And \before{D_m} \And 
			\beforeeq{E_1} \And \ldots \And \beforeeq{E_n} \And
			\aftereq{F_1} \And \ldots \And \aftereq{F_{m'}} \And
			\after{H_1} \And \ldots \And \after{H_{n'}}\\
\label{eq:almost_e} A &\to \epsilon,
\end{align}
\end{subequations}
with $a \in \Sigma$ and $A, B, C, D_i, E_i, F_i, H_i \in N$.
\end{lemma}

\begin{construction}\label{Construction:epsilon_elimination}
Let $G = (\Sigma, N, R, S)$ be a grammar with two-sided contexts,
with all rules of the form~(\ref{eq:rules_almost_normal_form}).
Consider the sets
$\textsc{Nullable}(G)$,
$\NullableLeftEps(G)$ and
$\NullableRightEps(G)$,
and
construct another grammar with two-sided contexts
$G' = (\Sigma, N, R', S)$,
with the following rules.
\begin{enumerate}
\item
All rules of the form~(\ref{eq:almost_a}) in $R$
are added to $R'$.

\item
Every rule of the form~(\ref{eq:almost_BC})
in $R$
is added to $R'$,
along with the following extra rules,
where a nullable nonterminal
is omitted
and the fact that it generates $\epsilon$
is
expressed by context operators.
\begin{align*}
A &\to B \And \beforeeq{J_1} \And \ldots \And \beforeeq{J_s} \And \after{K_1} \And \ldots \And \after{K_t},
\quad \text{for $(\{J_1, \ldots, J_s\}, C, \{K_1, \ldots, K_t\}) \in \textsc{Nullable}(G)$} \\
A &\to B \And \beforeeq{J_1} \And \ldots \And \beforeeq{J_s} \And \after{\epsilon},
\quad \text{for $(\{J_1, \ldots, J_s\}, C) \in \NullableRightEps(G)$ with $s \geqslant 1$} \\
A &\to C \And \before{J_1} \And \ldots \And \before{J_s} \And \aftereq{K_1} \And \ldots \And \aftereq{K_t},
\quad \text{for $(\{J_1, \ldots, J_s\}, B, \{K_1, \ldots, K_t\}) \in \textsc{Nullable}(G)$} \\
A &\to C \And \aftereq{K_1} \And \ldots \And \aftereq{K_t} \And \before{\epsilon},
\quad \text{for $(B, \{K_1, \ldots, K_t\}) \in \NullableLeftEps(G)$
with $t \geqslant 1$}
\end{align*}
In the first case,
$C$ defines
$\epsilon$
in
left contexts $J_i$ and
right contexts $K_i$,
and
this restriction is implemented by context operators in the new rule.
Since the left context of $C$
includes $B$,
extended context operators ($\beforeeq{J_i}$)
are used on the left,
whereas the right context operators are proper ($\after{K_i}$).

The second case
considers the possibility
of a nullable nonterminal $C$,
which defines $\epsilon$
in an empty right context.
This condition is
simulated
by the conjunct $\after{\epsilon}$
and extended left contexts $\beforeeq{J_i}$.

The two last rules handle symmetrical cases,
when the nonterminal $B$ defines
the empty string.

\item
Every rule of the form~(\ref{eq:almost_B_D_E_F_H})
is preserved in $R'$.
In the original grammar,
this rule~(\ref{eq:almost_B_D_E_F_H})
may generate strings in empty
contexts,
as long as
symbols in the context operators ($\before{D_i}$, $\after{H_i}$)
are nullable.

For any collection of pairs
$(D_1, V_1)$, \ldots, $(D_m, V_m) \in \NullableLeftEps(G)$,
with $m \geqslant 1$,
add the rule
\begin{equation*}
A \to
	B_1 \And \ldots \And B_k \And
	E_1 \And \ldots \And E_n \And
	\aftereq{K_1} \And \ldots \And \aftereq{K_t} \And
	\aftereq{F_1} \And \ldots \And \aftereq{F_{m'}} \And
	\after{H_1} \And \ldots \And \after{H_{n'}}
	\And \before{\epsilon},
\end{equation*}
where $\{K_1, \ldots, K_t\} = \bigcup_{i=1}^{m} V_i$.
Nonterminals $D_1$, \ldots, $D_m$
define $\epsilon$
in the right contexts given
in the set $\NullableLeftEps(G)$.
This is represented by
conjuncts $\before{\epsilon}$ and
$\aftereq{K_i}$.
Extended left contexts $\beforeeq{E_i}$
are replaced with base conjuncts $E_i$,
because in the empty left context
they have the same effect.

Symmetrically,
if
$(U_1, H_1)$, \ldots, $(U_{n'}, H_{n'}) \in \NullableRightEps(G)$,
with $n' \geqslant 1$,
then there is a rule
\begin{equation*}
A \to
	B_1 \And \ldots \And B_k \And
	F_1 \And \ldots \And F_{m'} \And
	\before{D_1} \And \ldots \And \before{D_m} \And
	\beforeeq{E_1} \And \ldots \And \beforeeq{E_n} \And
	\beforeeq{K_1} \And \ldots \And \beforeeq{K_t} \And
	\after{\epsilon},
\end{equation*}
where
$\{K_1, \ldots, K_t\} = \bigcup_{i=1}^{n'} U_i$.


Finally,
if
with $m$, $n' \geqslant 1$ and
$(D_1, V_1)$, \ldots, $(D_m, V_m) \in \NullableLeftEps(G)$,
$(U_1, H_1)$, \ldots, $(U_{n'}, H_{n'}) \in \NullableRightEps(G)$,
then the set $R'$ contains a rule
\begin{equation*}
A \to
	B_1 \And \ldots \And B_k \And
	E_1 \And \ldots \And E_n \And
	F_1 \And \ldots \And F_{m'} \And
	K_1 \And \ldots \And K_t \And
	\before{\epsilon} \And \after{\epsilon},
\end{equation*}
where $\{K_1, \ldots, K_t\} = \bigcup_{i=1}^{m} V_i \cup \bigcup_{j=1}^{n'} U_j$.
In this case,
both left and right contexts of a string are empty.
All the
symbols $D_i$ and $H_i$
define $\epsilon$ in the contexts
specified
in $\NullableLeftEps(G)$ and $\NullableRightEps(G)$.
These contexts apply to the entire string
and are explicitly stated as $K_1 \And \ldots \And K_t$
in the new rule.
The null contexts $\before{\epsilon}$, $\after{\epsilon}$
limit the applicability of this rule
to the whole string.
Again,
as in the two previous cases,
the base conjuncts are used
instead of extended context operators.
\end{enumerate}
\end{construction}

\begin{lemma}\label{epsilon_conjunct_elimination_lemma}
Let $G = (\Sigma, N, R, S)$ be a grammar with two-sided contexts.
Then the grammar $G' = (\Sigma, N', R', S)$
obtained by Construction~\ref{Construction:epsilon_elimination}
generates the language
$L(G') = L(G) \setminus \{\epsilon\}$.
\end{lemma}

The above construction eliminates the empty string
in all base conjuncts,
but the resulting grammar
may still contain null context specifications
($\before{\epsilon}$ and $\after{\epsilon}$),
which state that the current substring
is a prefix or a suffix of the whole string.
These operators are eliminated by the following simple transformation.
%
%
%
%
%
%
%
First, define a new nonterminal symbol $U$
that generates all non-empty strings in the empty left context.
This is done by the following three rules:
\begin{align*}
	U &\to U a
		&& \text{(for all $a \in \Sigma$)} \\
	U &\to a \And \beforeeq{X}
		&& \text{(for all $a \in \Sigma$)} \\
	X &\to a
		&& \text{(for all $a \in \Sigma$)}
\end{align*}
Another symbol $V$ generates all non-empty strings
in the empty right context;
it is defined by symmetric rules.
Then it remains to replace
left and right null context operators
($\before{\epsilon}$, $\after{\epsilon}$)
with $U$ and $V$, respectively.

The third stage of the transformation to the normal form
is removing the \emph{unit conjuncts}
in rules of the form $A \to B \And \ldots$
Already for conjunctive grammars~\cite{Conjunctive},
the only known transformation
involves substituting all rules for $B$
into all rules for $A$;
in the worst case, this results in an exponential blowup.
The same construction applies verbatim
to grammars with contexts.

This three-stage transformation proves the following theorem.

\begin{theorem}\label{thm_normal_form}
For each grammar with two-sided contexts $G=(\Sigma,N,R,S)$ 
there exists and can be effectively constructed 
a grammar with two-sided contexts $G'=(\Sigma,N',R',S)$
in the binary normal form,
such that $L(G)=L(G') \setminus \{\epsilon\}$.
\end{theorem}

\section{Parsing algorithm}\label{section_parsing_algorithm}

Let $G=(\Sigma,N,R,S)$ be a grammar with two-sided contexts
in the binary normal form, 
and let 
$w=a_1 \ldots a_n \in \Sigma^+$,
with $n \geqslant 1$ and $a_i \in \Sigma$,
be an input string to be parsed.
For every substring of $w$
delimited by two positions $i,j$,
with $0 \leqslant i < j \leqslant n$,
consider the set of nonterminal symbols
generating this substring.
\begin{equation*}
T_{i,j} = \makesetbig{A}{A \in N, \;
\stt{a_1 \ldots a_i}{a_{i+1} \ldots a_j}{a_{j+1} \ldots a_n} \in L_G(A)}
\end{equation*}
In particular, the whole string $w$ is in $L(G)$
	if and only if
$S \in T_{0,n}$.

\begin{figure}[t]
\begin{center}
\includegraphics[scale=0.8]{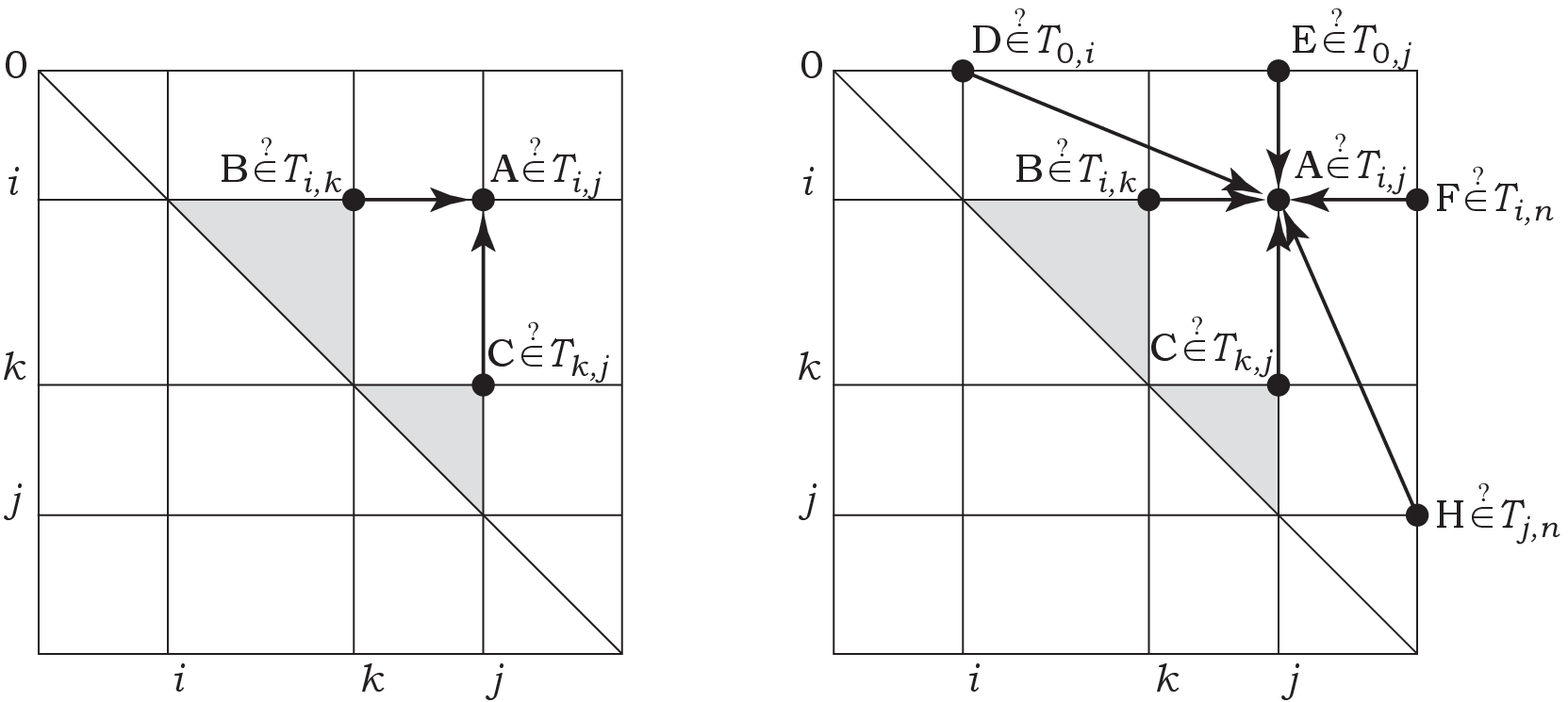}
\end{center}
\caption{How the membership of $A$ in $T_{i,j}$
	depends on other data,
	for rules (a) $A \to BC$
	and (b) $A \to BC \And \before{D} \And \beforeeq{E} \And \aftereq{F} \And \after{H}$.}
\label{f:dependencies_with_contexts}
\end{figure}

In ordinary context-free grammars,
a substring $a_{i+1} \ldots a_j$ is generated by $A$
if there is a rule $A \to BC$
and a partition of the substring
into $a_{i+1} \ldots a_k$ generated by $B$
and $a_{k+1} \ldots a_j$ generated by $C$,
as illustrated in Figure~\ref{f:dependencies_with_contexts}(left).
Accordingly, each set $T_{i,j}$ depends only on the sets $T_{i', j'}$
with $j'-i'<j-i$,
and hence all these sets may be constructed inductively,
beginning with shorter substrings and eventually reaching the set $T_{0,n}$:
this is the Cocke--Kasami--Younger parsing algorithm.
For conjunctive grammars,
all dependencies are the same,
and generally the same parsing algorithm applies~\cite{Conjunctive}.
In grammars with only left contexts,
each set $T_{i,j}$ additionally depends
on the sets $T_{0,i}$ and $T_{0,j}$
via the conjuncts of the form $\before{D}$ and $\beforeeq{E}$,
respectively,
which still allows constructing these sets
progressively for $j=1, \ldots, n$~\cite{grammars_with_contexts}.

The more complicated structure of logical dependencies
in grammars with two-sided contexts
is shown in Figure~\ref{f:dependencies_with_contexts}(right).
The following example demonstrates
how these dependencies may form circles.

\begin{example}
Consider the grammar with the rules
\begin{align*}
S &\to AB\\ 
A &\to a \And \after{B}\\ 
B &\to b \And \before{C}\\
C &\to a 
\end{align*}
and the input string $w=ab$.
It is immediately seen that $C \in T_{0,1}$.
From this, one can infer that $B \in T_{1,2}$,
and that knowledge can in turn be used to show that $A \in T_{0,1}$.
These data imply that $S \in T_{0,2}$.
Thus, 
none of the sets $T_{0,1}$ and $T_{1,2}$
can be fully constructed before approaching the other.
\end{example}

The proposed algorithm for constructing the sets $T_{i,j}$
works as follows.
At the first pass,
it makes all deductions
$\vdash_{G} \cpt{A}{a_1 \ldots a_i}{a_{i+1} \ldots a_j}{a_{j+1} \ldots a_n}$
that do not involve any contexts,
and accordingly puts $A$ to the corresponding $T_{i,j}$.
This pass progressively considers longer and longer substrings,
as done by the Cocke--Kasami--Younger algorithm for ordinary grammars.
During this first pass,
some symbols may be added to any sets $T_{0,j}$ and $T_{i,n}$,
and thus it becomes known that some contexts are true.
This triggers another pass over all entries $T_{i,j}$,
from shorter substrings to longer ones,
this time using the known true contexts in the deductions.
This pass may result in adding more elements to $T_{0,j}$ and $T_{i,n}$,
which will require yet another pass, and so on.
Since a new pass is needed
only if a new element is added to any of $2n-1$ subsets of $N$,
the total number of passes is at most $(2n-1) \cdot |N| + 1$.

These calculations are implemented in Algorithm~\ref{alg:parsing_cky},
which basically deduces all true statements
about all substrings of the input string.
For succinctness,
the algorithm uses the following notation
for multiple context operators.
For a set $\mathcal{X} = \{X_1, \ldots, X_\ell\}$,
with $X_i \in N$,
and for 
an operator $Q \in \{\before{}, \beforeeq{}, \aftereq{}, \after{}\}$,
denote
$Q \mathcal{X} := Q X_1 \And \ldots \And Q X_\ell$.

\begin{algorithm}
\begin{myalgorithm}\label{alg:parsing_cky}
Let $G=(\Sigma,N,R,S)$
be a grammar with contexts in the binary normal form.
Let $w=a_1 \ldots a_n \in \Sigma^+$ (with $n \geqslant 1$ and $a_i \in \Sigma$) 
be the input string. 
Let $T_{i,j}$ with $0 \leqslant i < j \leqslant n$
be variables, each representing a subset of $N$,
and let $T_{i,j}=\emptyset$ be their initial values.

\begin{algorithmic}[1]
\WHILE{any of $T_{0,j}$ ($1 \leqslant j \leqslant n$)
or $T_{i,n}$ ($1 \leqslant i < n$) change}
\label{cubic_time_parsing__loop_while_changes}
	\FOR{$j=1,\ldots,n$}
	\label{cubic_time_parsing__loop_j}
		\FORALL{$A \to a \And
		\before{\mathcal{D}} \And
		\beforeeq{\mathcal{E}} \And
		\aftereq{\mathcal{F}} \And
		\after{\mathcal{H}} \in R$}
		\label{cubic_time_parsing__a_j}
			\IF{$a_j=a$ $\land$ $\mathcal{D} \subseteq T_{0,j-1}$ $\land$
				$\mathcal{E} \subseteq T_{0,j}$ $\land$
				$\mathcal{F} \subseteq T_{j,n}$ $\land$
				$\mathcal{H} \subseteq T_{i,n}$}
				\STATE $T_{j-1,j} = T_{j-1,j} \cup \{A\}$
				\label{cubic_time_parsing__T_jminus1_j}
			\ENDIF
		\ENDFOR
		\FOR{$i=j-2$ to $0$}
		\label{cubic_time_parsing__loop_i}
			\STATE let $P = \emptyset$ ($P \subseteq N \times N$)
			\FOR{$k=i+1$ to $j-1$}
			\label{cubic_time_parsing__loop_k}
				\STATE $P = P \cup (T_{i,k} \times T_{k,j})$
				\label{cubic_time_parsing__cartesian_product}
			\ENDFOR
			\FORALL{$A \to B_1 C_1 \And \ldots \And B_m C_m \And
				\before{\mathcal{D}} \And
				\beforeeq{\mathcal{E}} \And
				\aftereq{\mathcal{F}} \And
				\after{\mathcal{H}} \in R$}
			\label{cubic_time_parsing__loop_T_i_j}
				\IF{$(B_1,C_1), \ldots, (B_m, C_m) \in P$ $\, \land \,$
					$\mathcal{D} \subseteq T_{0,i}$ $\, \land \,$
					$\mathcal{E} \subseteq T_{0,j}$ $\, \land \,$
					$\mathcal{F} \subseteq T_{j,n}$ $\, \land \,$
					$\mathcal{H} \subseteq T_{i,n}$}
					\STATE $T_{i,j} = T_{i,j} \cup \{A\}$
					\label{cubic_time_parsing__T_i_j}
				\ENDIF
			\ENDFOR 
		\ENDFOR 
	\ENDFOR
\ENDWHILE
\STATE{accept if and only if $S \in T_{0,n}$}
\end{algorithmic}
\end{myalgorithm}
\end{algorithm}

\begin{theorem}\label{cubic_time_parsing_theorem}
For every grammar with two-sided contexts $G$
in the binary normal form,
Algorithm~\ref{alg:parsing_cky},
given an input string $w=a_1 \ldots a_n$,
constructs the sets $T_{i,j}$
and determines the membership of $w$ in $L(G)$,
and does so in time $\mathcal{O}(|G|^2 \cdot n^4)$,
using space $\mathcal{O}(|G| \cdot n^2)$.
\end{theorem}

While this paper was under preparation,
Rabkin~\cite{Rabkin} developed
a more efficient and more sophisticated parsing algorithm
for grammars with two-sided contexts,
with the running time $\mathcal{O}(|G| \cdot n^3)$,
using space $\mathcal{O}(|G| \cdot n^2)$.
Like Algorithm~\ref{alg:parsing_cky},
Rabkin's algorithm works by proving all true statements
about the substrings of the given string,
but does so using the superior method of Dowling and Gallier~\cite{DowlingGallier}.
Nevertheless, Algorithm~\ref{alg:parsing_cky} retains some value
as the elementary parsing method
for grammars with two-sided contexts---%
just like the Cocke--Kasami--Younger algorithm for ordinary grammars
remains useful, in spite of the asymptotically superior Valiant's algorithm~\cite{Valiant}.

\section{Conclusion}

This paper has developed a formal representation
for the idea of phrase-structure rules applicable in a context,
featured in the early work of Chomsky~\cite{Chomsky}.
This idea did not receive adequate treatment at the time,
due to the unsuitable string-rewriting approach.
The logical approach,
adapted from Rounds~\cite{Rounds} and his predecessors,
brings it to life.

There are many theoretical questions to research about the new model:
for instance, one can study
the limitations of their expressive power,
their closure properties,
efficient parsing algorithms
and subfamilies that admit more efficient parsing.
Another possibility for further studies
is investigating Boolean and stochastic variants
of grammars with contexts,
following the recent related work~\cite{EsikKuich,Kountouriotis_et_al,Ziervogel}.

On a broader scope,
there must have been other good ideas
in the theory of formal grammars
that were inadequately formalized before.
They may be worth being re-investigated using the logical approach.


\end{document}
